# Natural Language Aggregate Query over RDF Data

Xin Hu, Yingting Yao, Luting Ye, Depeng Dang*

*College of Information Science and Technology, Beijing Normal University, Beijing 100875, China*
*Corresponding author. Tel: +86 13121915369 E-mail address: ddepeng@bnu.edu.cn

**ABSTRACT**:

Natural language question/answering over RDF data has received widespread attention. Although there have been several studies that have dealt with a small number of aggregate queries, they have many restrictions (i.e., interactive information, controlled question or query template). Thus far, there has been no natural language querying mechanism that can process general aggregate queries over RDF data. Therefore, we propose a framework called NLAQ (Natural Language Aggregate Query). First, we propose a novel algorithm to automatically understand a user's query intention, which mainly contains semantic relations and aggregations. Second, to build a better bridge between the query intention and RDF data, we propose an extended paraphrase dictionary *ED* to obtain more candidate mappings for semantic relations, and we introduce a predicate-type adjacent set *PT* to filter out inappropriate candidate mapping combinations in semantic relations and basic graph patterns. Third, we design a suitable translation plan for each aggregate category and effectively distinguish whether an aggregate item is numeric or not, which will greatly affect the aggregate result. Finally, we conduct extensive experiments over real datasets (QALD benchmark and DBpedia), and the experimental results demonstrate that our solution is effective.

## 1. Introduction

As more and more data are available on the web, academics and industry researchers must invest much more in bold strategies that can achieve natural language searching and answering [1]. RDF (Resource Description Framework) has been widely used as a W3C standard to describe data in the Semantic Web. Thus, natural language question/answering (Q/A) over RDF data has received widespread attention [31, 32, 33, 34]. Although these methods are easy to use and can produce interesting results, they do not accommodate even simple aggregate queries, such as "*How many books by Kerouac were published by Viking Press?*"

Few works have dealt with a small number of aggregate queries over RDF data [29, 14, 15, 28], and users cannot access RDF data conveniently. Some of these works constructed an interactive interface [14, 15], which requires users to fill out or choose aggregate items and aggregate categories. The input of Squall2sparql is a controlled English question [28], and users need to specify the precise entities and predicates (denoted by URIs) in the question. TBSL [29] is a template-based approach and does not require users to do something extra, but the query templates in TBSL are fixed and need to be constructed by analyzing a huge set of candidate queries. In conclusion, they answer aggregate queries over RDF data with too many restrictions and can only deal with a small number of aggregate queries. The main reason for this is that identifying and transforming aggregates are really difficult issues.

In addition to this, there are two stages that need to be improved in RDF Q/A systems: *query understanding* and *mapping*. In the first stage, existing researches [28,29,31,32,33,34] regarding the identification of semantic relations totally depend on the verb phrase in the query and paraphrase dictionary *D*, which records the semantic equivalence between verb phrases and predicates. The basic idea is to find two associated arguments of *rel* in the query according to linguistic rules, where *rel* is also a verb phrase in *D*. Then, the verb phrase *rel*, together with two associated arguments, forms a semantic relation <*arg1, rel, arg2*>. However, there is a major disadvantage in this method. For Query1, "*How many books by[1] Kerouac were published by Viking Press?*," the verb phrase "*published*" is most likely to be found in *D*, while the non-verb phrase "*by[1]*" is not. Therefore, existing methods can identify the triple <*Kerouac, published, Viking Press*> and overlook the triple <*books, by, Kerouac*>.

In the second stage, existing researches [28,29,31,32,33,34] have not been able to obtain more candidate mappings for semantic relations and effectively filter out inappropriate mappings when the mappings have the same (or approximate) similarity score. Their basic idea is to strictly map the verb phrase *rel* and arguments *arg1/arg2* to the candidate predicate and entity/type, respectively, and then some sets of candidate mappings with high similarity scores are selected. On the one hand, strictly mapping can improve the accuracy of mapping for a query that has no ambiguity. However, natural language has a wide range of ambiguity, and strict mapping will reduce the number of candidate mappings of triples and make most queries unanswerable (see the example in section 5.2.1). On the other hand, after mapping, existing methods depend on similarity scores alone to select candidate mappings and will produce many irrelevant sets. For example, in the semantic relation <*books, published, Viking_Press*>, the *rel* "*published*" has been mapped to the predicates "*dbo:publisher*," "*dbo:publishedIn*" and "*dbp:publishDate*," as shown in Table 3. All of these predicates have the same similarity score of 0.6, with only "*dbo:publisher*" being relevant, as shown in Fig. 1. For this situation, existing researches cannot solve it.

Therefore, we propose a framework called NLAQ (Natural Language Aggregate Query) that can process general natural language aggregate queries and improve the capability of natural language question/answering over RDF data. We make the following contributions in this paper:

1. We perform a first step toward processing natural language aggregate queries over RDF data via automated identification and transformation of aggregations rather than restrictions, such as controlled English question, interactive information and query template.
2. During query understanding, we propose algorithm AIII to automatically identify intention interpretations (i.e., semantic relations, question items and aggregations) from the natural language aggregate query. This can overcome the shortcomings of existing methods in that they neglect some semantic relations and cannot identify aggregations.
3. During the mapping stage, on the one hand, to get more candidate mappings, we propose the extended paraphrase dictionary *ED*, which appends the semantic equivalence between arguments of the semantic relation and predicates to the original paraphrase dictionary *D*. On the other hand, we propose the predicate-type adjacent set *PT* and the subset *PP* of *PT* to filter out inappropriate mapping combinations in semantic relations and basic graph patterns, respectively.
4. For a variety of aggregate categories, we design a suitable translation plan for each aggregate category and effectively distinguish whether the aggregate item is numeric or not, which will greatly affect the aggregate result.

## 2. Related work

RDF is a W3C standard to represent information that has currently gained much attention in real applications, such as the Semantic Web. Previous works have typically studied the problem of data models (for example, triple store [2,3,4], column store [5,6,7], property tables [8,9] and graphs [10,11]) and the efficiency of SPARQL query answering (for example, RDF-3X [3], Hexastore [4], C-Store [5], MonetDB [6], and gStore [12,13]).

As the stores and queries mature, expanding the queries is beginning to attract attention. To issue a standard query, users must know the schema of data and the syntax of a standard query. While expressive and powerful, standard query language (i.e., SPARQL/SQL/XQuery) is too difficult for users without technical training. Many researchers have provided querying mechanisms that can be used by ordinary users to explore complex databases.

*2.1 Non-natural language question/answering*

*Keyword*. The first category is that users express query intentions with various simple keywords. Keyword search has already been studied in the context of relational databases [19,20], XML documents [22,23] and RDF data [21,24]. Among them, PowerQ [19] and SQAK [20] can process aggregate queries over relational databases via simple keywords, but PowerQ needs an interactive interface and SQAK strictly limits the location of keywords.

*Interactive interface*. The second category is to construct an interactive interface that employs feedback and clarification dialogs to resolve ambiguities and improve the domain lexicon with the help of users [14,15,17,18,19,27, 45]. User feedback is used to enrich the semantic matching process by allowing manual query-vocabulary mapping. The interaction techniques require users to select a number of options from lists or write words in blank squares. Among them, [14,15,45] construct interactive interfaces for RDF data, [17,18,19] for relational databases and [27] for XML databases.

On the one hand, natural language queries have stronger expressive power than keyword queries and can express diverse queries. On the other hand, although some interactive interfaces can process various aggregate queries, users need to continually provide much of the interactive information. In contrast, we believe that natural language queries are superior to expended queries.

*2.2 Natural language question/answering over relational/XML databases*

Roy et al. [25] introduced a principled approach to provide explanations for answers to SQL queries based on intervention: removal of tuples from the database that significantly affect the query answers. Bais et al. [49] presented the architecture and implementation of a generic natural language interface based on a machine learning approach for a relational database. Alghamdi et al. [50] proposed a novel approach for building a Natural Language Interface to a Relational Database (NLI-RDB) using Conversational Agent (CA), Information Extraction (IE) and Object

Relational Mapping (ORM) frameworks. Joseph et al. [51] and Li et al. [52,53] propose a system that can accept English language sentences and then translate them into an XQuery expression.

Different data types will lead to different processing techniques of natural languages and aggregations. NLAQ translates natural language queries into SPARQL rather than SQL/XQuery; thus, we cannot borrow previous techniques and have to design our own method.

*2.3 Natural language question/answering with aggregation over RDF data*

Based on controlled natural languages, the approaches in [28,46] consider a well-defined restricted subset of natural language that can be unambiguously interpreted by a given system. However, its input is controlled English questions rather than a truly natural language question. TBSL [29] is a template-based approach. It constructs some templates based on a linguistic analysis of the input question. Then, these templates are instantiated by matching the natural language expressions occurring in the question with elements from the queried dataset. However, the constructed query templates are too fixed, and a huge set of candidate queries needs to be considered; thus, the diversity of questions that can be answered is limited. To tackle this problem, Zheng et al. [30] studied how to generate templates automatically, but aggregate queries are still a roadblock to TBSL.

Different from [28,46,29,30], which can only answer a small number of aggregate queries, users can access RDF data conveniently, namely, NLAQ can answer natural language aggregate queries without the above restrictions (i.e., controlled English language, query template). Moreover, we build a framework that can process general natural language aggregate queries so that our method can answer most aggregate queries.

*2.4 Natural language question/answering without aggregation over RDF data*

Zou et al. [31] proposed an entire-graph data-driven framework to answer natural language questions over RDF graphs and push down the disambiguation into the query evaluation stage. Amsterdamer et al. [32] studied the problem of translating natural language questions that involve general and individual knowledge into formal queries. Fader et al. [33] introduced a novel open Q/A system that is the first to leverage both curated and extracted knowledge. Yahya et al. [34,35,36,38] analyzed questions and mapped verbal phrases to relations and noun phrases to either individual entities or semantic classes. Lopez et al. [37] proposed a system that takes queries expressed in natural language and an ontology as input and returns answers drawn from the available semantic markup. Liu et al. [47] proposed a method for constructing directed acyclic graphs and triples, and the parsing for the modifier constraint greatly improves the conversion efficiency. Rozinajová et al. [48] proposed a method based on a sentence structure, utilizing dependencies between the words in user queries.

Different from most existing RDF Q/A systems [31,32,33,34,35,36,37,38,47,48], which ignore aggregate queries, we can answer the natural language aggregate queries and improve the capability of RDF Q/A regarding the non-aggregated part in queries from two aspects: *query understanding* and *mapping*.

*Query understanding*. Zou et al [31] first applied the Stanford Parser to query $N$ to obtain the dependency tree $Y$ of $N$, and they then extracted the semantic relations from $Y$ based on the paraphrase dictionary $D$, which records the semantic equivalence between relation phrases and predicates. However, if some semantic relations in query $N$ do not contain relation phrases in $D$, the method cannot identify these semantic relations. This is similar in other research studies [32,33,34,35,36,37,38,47,48]: if the relation phrase between the subject and object in a semantic relation is not a verb phrase in the query (see the example in section 1), the method in these research studies cannot identify the semantic relation. In contrast, we automatically identify intention interpretations (semantic relations and aggregates) without requiring relation phrases, and we can identify more semantic relations that are often overlooked by these research studies and aggregate information.

*Mapping*. Almost all existing studies have phrase mapping. We do not change the method of mapping and just improve the effectiveness of mapping via the extended paraphrase dictionary $ED$, which can be used to get more semantic relation mappings, and the proposed predicate-type adjacent set $PT$, which can be used to delete many inappropriate combinations.

Besides the above literature, there are some natural language question/answering systems that pay attention to many other interesting research directions. Sun et al. [26], Balakrishna et al. [54] and Tatu et al. [55] mined answers from integrated structured data and unstructured data. El-Ansari et al. [56] presented a Question Answering system that combines multiple knowledge bases. Freitas et al. [16] proposed and evaluated the suitability of the distributional-compositional semantics model applied to the construction of a question answering system for linked data. Mervin et al. [57] presented how sentences in the English language can be represented as knowledge patterns by means of RDF. Shekarpour et al. [58] proposed a new method for automatic rewriting of input queries on graph-structured RDF knowledge bases. Amsterdamer et al. [59] developed NL2CM, a prototype system that translates natural language (NL) questions into well-formed crowd-mining queries. Dubey et al. [60] proposed AskNow based on a novel intermediary canonical syntactic form. Scholten et al. [61] and Hamon et al. [62] proposed systems that query biomedical linked data with natural language questions.

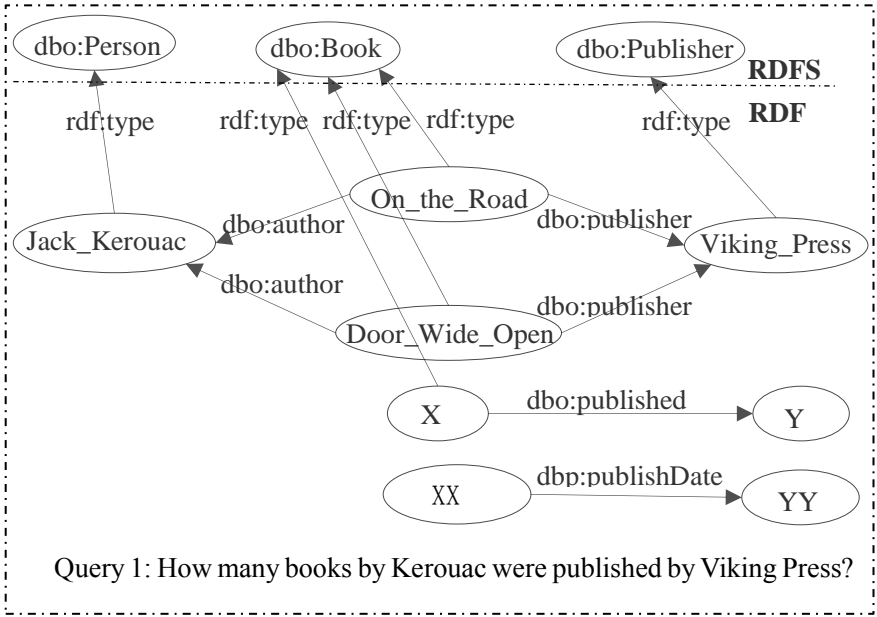

Fig. 1. RDF(S) data and sample queries.

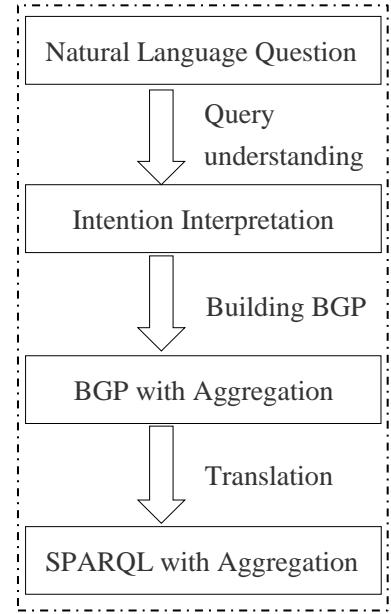

Fig. 2. Architecture of NLAQ

## 3. Overview

NLAQ solves the problem of ordinary users processing natural language aggregate queries over RDF data. Fig. 1 shows RDF data and an example query. Fig. 2 provides an overview of the NLAQ architecture.

There are three key stages in this paper: 1) how to represent the questioner's query intention by analyzing the query $N$ (*Query Understanding*); 2) how to correctly express the query intention using the information of the RDF repository (*Building Basic Graph Pattern-BGP*); and 3) how to translate BGP to SPARQL with aggregation (*Translation*).

### 3.1 Query Understanding

We automatically extract the intention interpretation (Definition 2) implied by the query $N$. The intention interpretation contains semantic relations, question items, aggregate items and aggregate categories. In contrast, existing research studies can identify semantic relations that contain a verb phrase and question items. They cannot identify semantic relations that do not contain a verb phrase, aggregate items and aggregate categories.

**DEFINITION 1**. (Semantic Relation). A semantic relation is a triple denoted as $R<arg1, rel, arg2>$, where *rel* is a relation phrase and *arg*1 and *arg*2 are two arguments.

**Example 1**. For query 1 in Fig. 1, <*books, published, Viking_Press*> is a semantic relation, in which "*published*" is the relation phrase *rel* and "*books*" and "*Viking_Press*" are two associated arguments *arg1* and *arg2*, respectively. We can also find another semantic relation <*books, by, Kerouac*> in query 1.

**DEFINITION 2**. (Intention Interpretation). An intention interpretation is denoted as $I= \{S, Q, A\}$, where
$S= \{R_i | R_i$ is the *i*-th semantic relation$\}$.
$Q= \{q_i | q_i$ is the *i*-th question item$\}$.
$A= \{<a_i, c_i> \mid a_i, c_i$ are the *i*-th aggregate item and aggregate category, respectively$\}$.
For Query1, $I=\{S=\{<books, by, Kerouac>,<books, published, Viking\_Press>\},Q=\{books\},A=\{<books,count>\}\}$.

### 3.2 Building the Basic Graph Pattern

#### 3.2.1 Semantic Relation Mapping

To correctly express the query intention using the information of the RDF repository, we introduce two important stages: *phrase mapping* and *semantic relation mapping*.

*Phrase Mapping*. The technology of phrase mapping has become very mature, and we just propose the extended paraphrase dictionary *ED*, which makes arguments of the semantic relation that can be mapped to predicates so that we can obtain more and better candidate mappings than existing researches.

*Semantic Relation Mapping.* Semantic relation mapping is still a difficult challenge. We construct the predicate-type adjacent set *PT* (DEFINITION 3) to improve semantic relation mapping. We use the set *PT* to filter or recommend candidate mappings of semantic relations and sometimes adjust the positions of arguments for some specific mappings; then, we can obtain better semantic relation mappings than existing research studies.

**DEFINITION 3**. (Predicate-type Adjacent set, *PT*). $PT=\{(T_i - P_k - T_j) \text{ and } (P_i - T_k - P_j)\}$, where $(T_i - P_k - T_j)$ represents that *m* and *n* are of type $T_i$ and $T_j$, respectively, and *m* and *n* come from a triple $<m, P_k, n>$. $(P_i - T_k - P_j)$ represents that *y* is of type $T_k$, and *y* comes from two connected triples $(x, P_i, y), (y, P_j, z)$.

**Example 2**. We can generate the *PT* sets that come from the RDF data as shown in Fig. 1: *PT*={(*dbo:Book-dbo:author-dbo:Person*) (*dbo:Book-dbo:publisher-dbo:Publisher*), (*Ø-dbo:Book-dbo:publisher/dbo:author*), (*Ø-dbo:Book-dbo:publishedIn*)}.

**DEFINITION 4**. (The Score of One Semantic Relation Mapping, *s(RM)*). *RM* represents one of the mapping of the semantic relation *R*, and *s(RM)* represents the score of *RM*. *s(RM)* is the total mapping score of *arg*1, *rel* and *arg*2 because one inaccurate component mapping has little impact on the overall *RM*.

$$s(RM) = s(M_{arg1}) + s(M_{rel}) + s(M_{arg2}),$$

where $M_{arg1}$ represents the mapping of *arg*1 and $s(M_{arg1})$ represents the score of $M_{arg1}$ and comes from *ED*. Furthermore, if *arg1* or *arg2* corresponds to a constant, its mapping score is 1.

**Example 3**. The semantic relation <*books, published, Viking_Press*> has one mapping <*dbo:Book*-1.0, *dbo:publisher*-0.6, *Viking_Press*-constant> so that we can get the score *s(RM)*=1.0+0.6+1.0=2.6.

### 3.2.2 Building the Basic Graph Pattern

One query may contain multiple semantic relations such that we need to combine several semantic relation mappings and filter out inappropriate combinations by using the predicate-predicate adjacent set *PP* derived from *PT*. Then, we will select the top *k* highest-scoring basic graph patterns.

**DEFINITION 5**. (The Score of One Basic Graph Pattern, *s(BGP)*). A query may have many candidate basic graph patterns (BGPs), and a group of mappings of all semantic relations are collected together to form a BGP. *s(BGP)* is the product of the score of all semantic relation mappings in a BGP because one inaccurate semantic relation mapping has a major impact on the overall BGP.

$$s(BGP) = \prod_{i=1}^{n} s(RM_i),$$

where *n* represents the number of semantic relations in BGP and $s(RM_i)$ represents the score of the *i*-th semantic relation mapping $RM_i$ in BGP.

**Example 4**. Query 1 in Fig. 1 has two triples: <*books, by, Kerouac*> and <*Kerouac, published, Viking Press*>. One BGP of the query is a group of mappings {<*dbo:Book, ?X, Kerouac*>-2.0, <*dbo:Book, dbo:publishedIn, Viking_Press*>-2.6}, and we can find that its score is *s(BGP)*=2.0*2.6=5.2.

### 3.3 Translation

Finally, we need to translate the basic graph pattern and aggregation into an executable SPARQL statement with aggregation. Due to the complexity of aggregation, we divide it into various categories and then carry out target translation. Given the diversity of basic graph patterns, the SPARQL statement of aggregation may be different for each basic graph pattern.

## 4. Query understanding

### 4.1 Dependency Structure

Some NLP (Natural Language Processing) literature suggests that the dependency structure is more stable for the relation extraction [39], and the Stanford parser (http://nlp.stanford.edu: 8080/parser/) is a very good tool to get the dependency structure. Therefore, we apply it to obtain the dependency structure from the query. Fig. 3 shows the dependency structure for query 1.

```
advmod(many-2, How-1)
amod(books-3, many-2)
nsubjpass(published-7, books-3)
case(Kerouac-5, by-4)
nmod:by(books-3, Kerouac-5)
auxpass(published-7, were-6)
root(ROOT-0, published-7)
case(Press-10, by-8)
compound(Press-10, Viking-9)
nmod:by(published-7, Press-10)
```

```
I={S={<books,by,Kerouac>,
      <books,published,Viking_Press>},
   Q={books}
   A={<books, count>}}.
```

Fig. 3. Dependency structure from the Stanford parser      Fig. 4. Intention interpretation for Query 1

*4.2 Categories of Dependency Structure and Rules of Combination*

*Categories of Dependency Structure*. There are some important dependency structures that we can use to produce intention interpretations and that can be divided into six categories, as shown in Table 1.

Table 1. Categories of dependency structures

| Category | Dependency structure | Intention |
|---|---|---|
| $\delta_{subject-like}$ | subj,nsubj,nsubjpass,csubj,csubjpass,xsubj | S |
| $\delta_{object-like}$ | obj, pobj, dobj, iobj | |
| $\delta_{s\_or\_o-like}$ | acl, nmod | |
| $\delta_{question-like}$ | amod,det,dobj,nsubj | Q |
| $\delta_{aggregation-like}$ | amod, nwe, nummod, nmod | A |
| $\delta_{constant-like}$ | compound | constant |

*Rules of Combination*. If the constant contains more than one word, we need to combine these words and rely on $\delta_{constant-like}$. Then, we will map these dependency structures to intention interpretations by the following rules:

1) $R(s,p,o)=f(\delta_{subject-like} \wedge \delta_{object-like})$
2) $R(s,p,o)=f((\delta_{subject-like}/\delta_{object-like}) \wedge \delta_{s\_or\_o-like})$
3) $R(s,p,o)=f(\delta_{s\_or\_o-like})$
4) $Q=f(\delta_{question-like})$
5) $A=f(\delta_{aggregation-like})$

Rule 1 means that we can get some semantic relations *R* by composing the dependency structure set $\delta_{subject-like}$ and $\delta_{object-like}$. Similarly, we can get other *R*, *Q* and *A* from the other dependency structures.

*4.3 Identify Intention Interpretations from Dependency Structures*

During this stage, existing methods will overlook semantic relations that do not contain a verb phrase, and they identify aggregation by restrictions (i.e., interactive information, controlled question or query template).

**Example 5**. For Query 1, existing methods can identify the triple <*Kerouac, published, Viking Press*> but overlook the triple <*books, by, Kerouac*>. Furthermore, these methods almost cannot identify the aggregate item *books* and aggregate category COUNT automatically.

To better identify semantic relations and aggregations, we propose an algorithm called AIII (Automatically Identify Intention Interpretation). The basic idea is to find important dependency structures from the result of the Stanford Parser and then analyze and combine these important dependency structures to produce the intention interpretation *I*. Fig. 4 shows the intention interpretation of query 1.

**Algorithm 1** AIII(Automatic Identify Intention Interpretation)
**Require**: **Input**: Natural language question *N*
        **Output**: Intention interpretation *I*
1: *D*=Stanford_Parser(*N*)
2: δ=Filter_divide_important_dependency(*D*)
3: *C*=Composit_constant($\delta_{constant-like}$)
4: δ=Update(δ,*C*)

5: $S$=Combine($\delta_{subject-like}$, $\delta_{oject-like}$)
6: $S$=$S$+Combine($\delta_{s\_or\_o-like}$, $\delta_{subject-like}$+$\delta_{object-like}$)
7: $S$=$S$+rest($\delta_{s\_or\_o-like}$)
8: $Q$=Get_question($\delta_{question-like}$)
9: $A$=Get_aggregation($\delta_{aggregation-like}$, $Q$)
10: $S$=$S$+$A$
11: $I$=Together($S,Q,A$)

Firstly, we get the important dependency structure set $\delta$ (line 1) and divide the set contents into six categories (line 2). Then, we get the constant from $\delta_{constant-like}$ (line 3) while it is not empty and update all constants in the set $\delta$ (line 4).

**Example 6**. The constant "*Viking Press*" comes from the dependency structures "*compound (Press-10, Viking-9)*" in $\delta_{constant-like}$, while the constant "*Kerouac*" has nothing to do.

Secondly, we generate semantic relations by combine two dependency structures in $\delta_{subject-like}$ and $\delta_{object-like}$ if the relation phrases *rel* of two dependency structures are the same phrases (line 5). Similarly, we get a new sematic relation if the dependency structures in $\delta_{s\_or\_o-like}$ can be combined with dependency structures in $\delta_{subject-like}$ or $\delta_{object-like}$ (line 6). Finally, we transform the rest of the dependency structures in $\delta_{s\_or\_o-like}$ to semantic relations.

**Example 7**. We get the semantic relation $R_1$<*books, published, Viking_Press*> by rule 2, which combines the dependency structures "*nsubjpass(published-7, books-3)*" and "*nmod:by(published-7, Press-10).*" We get the semantic relation $R_2$<*books,by,Kerouac*> from the dependency structure "*nmod:by(books-3, Kerouac-5)*" by rule 3.

Thirdly, to get the question item (line 8), we can divide the query into two cases as follows. 1) The question item is obvious, e.g., the question item is *yes/no* for the query "*Do…/Does…/Is…/……,*" and the question item is *time/place/person* for the query "*when…/where… /who….*" 2) The question item is not obvious, e.g., "*Which…/In which…/ What…/For what…/How many…/How many official languages…/List …/ Give me …./Show me…/…….*" However, through our research and analysis, for a query that contains "*which,*" we can get the question item from the dependency structure "*det,*" denoted as $\sigma_{which}$={*det*}. In the same manner, we can get the question item from $\sigma_{what}$= {*nsubj*}, $\sigma_{how\_many}$={*amod*}. In addition, for other queries such as "*List …/ Give me …./Show me…/……,*" we can get the question item from $\sigma_{others}$= {*dobj*}.

**Example 8**. Because the type of question is "*How many…,*" we get the question item $Q$={*books*} from "*amod(books-3, many-2)*" by rule 4.

Finally, to get the aggregate item and aggregate category (line 9), we can divide the query into two cases as follows. 1) The question item is also an aggregate item, such as "*How many/What's amount of/….,*" and we can get the aggregate category (i.e., COUNT/SUM) from the method of raising the question. 2) The question item can be obtained from $\delta_{aggregation-like}$. We get the aggregate category MAX/MIN/AVG/…, such as "*most, first, second, highest, average……*" from {*amod*}. In the same manner, we can get the aggregate category >/< from {*nwe, nummod, nmod*}, and so on.

**Example 9**. We get the aggregation "*A*={<*books, count*>}" from "*amod(books-3, many-2)*" by rule 5.

In addition, sometimes the word contains not only the aggregation but also predicates, and we need to add the word into the semantic relation *S* (line 10).

**Example 10**. For the query "*What is the largest city in Australia?*," we can get the intention interpretation *I*={*S*={<*city, in, Australia*>}, *Q*={*city*}, *A*={*city, largest*}} before line 10 and *I*={*S*={<*city, in, Australia*>**<*city, largest, #*>**}, *Q*={*city*}, *A*={*city, largest*}} after line 10. Consider two cases. 1) There is a triple *t1*=<*dbr:Australia, dbo:largestCity, dbr:Sydney*> in the RDF data. 2) There are triples *t2*=<*dbr:Sydney, dbo:populationTotal, xxx*> and *t3*=<*dbr:Australia, dbo:city, dbr:Sydney* > in the RDF data. For the second case, we have no semantic relation that can be mapped to *t2*. To solve this problem, we will add a new semantic relation (i.e., <*city, largest, #*>) to *S*. For the first case, after mapping the semantic relation <*city, in, Australia*> to *t1*, we found that the predicate "*dbo:largestCity*" contains an aggregation, and then we will delete the aggregation *A* and the semantic relation <*city, largest, #*> from *I*.

*4.4 Improving the Intention Interpretation*

Furthermore, if the semantic relation set *S* does not satisfy the following condition, we will provide an alternative possible semantic relation set *S*:

● If *arg1* is constant, *arg2* is a determined value and cannot be an aggregate item with an aggregate category > or < unless the query is a judgment sentence. We will replace the aggregate item with another argument that is the nearest argument to *arg1*.

**Example 11**. For the query "*Give me cities in New Jersey with more than 100000 inhabitants,*" due to the incorrect dependency structure "*nmod:with(Jersey-7, inhabitants-12)*" resulting from the Stanford parser, we get the incorrect intention interpretation *I*={ S=<*cities, in, New_Jersey*>, <~~New_Jersey, with, inhabitants~~>, *Q*={*cities*}, *A*={*inhabitants*,>100000}}. Thus, we replace <~~New_Jersey, with, inhabitants~~> with <*cities, with, inhabitants*> and get the new intention interpretation *I*={S=<*cities, in, New_Jersey*>, <*cities, with, inhabitants*>, *Q*={*cities*}, *A*={*inhabitants*,>100000}}.

## 5. Building Basic Graph Pattern

*5.1 Offline*

Different from existing research, the extended paraphrase dictionary *ED* is not used during query understanding. It will be used together with *PT* during phrase mapping.

*5.1.1 Extended Paraphrase Dictionary (ED)*
To improve the mapping between the semantic relation and RDF data, we propose the extended paraphrase dictionary *ED*. On the one hand, we keep the content of the paraphrase dictionary *D*, which records the semantic equivalence between *verb phrases* and *predicates*, *arguments* and *types*, as well as existing research studies [28,29,31,32,33,34]. On the other hand, we add the semantic equivalence between *arguments* and *predicates*. The method we used to get the extended paraphrase dictionary *ED* is not discussed, as it is the same method used in the related research studies [41,42,43,44] to get the dictionary *D*.

**Example 12**. The *ED* records the semantic equivalence between the *rel* "*published*" and the predicate "*dbo:publisher*," the argument "*books*" and the type "*dbo:Book*", which are also recorded in *D*. Furthermore, *ED* also records the semantic equivalence between the argument "*books*" with the predicate "*dbo:awardedBook*" as shown in Table 2.

Table 2. Extended paraphrase dictionary *ED*

| Phrases | Similar Semantic | Probability |
|---|---|---|
| *"published"* | *dbo:publisher* | 1.0 |
| *"books"* | *dbo:Book* | 1.0 |
| **"books"** | **dbo:awardedBook** | **0.5** |
| …… | …… | …… |

*5.1.2 Predicate-type Adjacent Set (PT)*
The paraphrase dictionary *ED* can improve the mapping between the semantic relation and RDF data. However, when the volume of data is very large, many phrases will have too many similar semantic predicates or types. Therefore, we build a predicate-type adjacent set *PT* (DEFINITION 3) to filter out inappropriate predicate mappings or type mappings in semantic relations (see example 15 in section 5.2.2). It also can provide some candidate mappings when the number of mappings is small due to spelling errors in the query (see example 17 in section 5.2.2). The method of getting the *PT* set is simple and only needs to execute a few SPARQL statements, as shown in Fig. 5.

```
SELECT    ?predicate1 ?Type1 ?predicate2
WHERE {optional{?s1   ?predicate1    ?s2}
         optional{?s2   ?predicate2    ?s3}.
                ?s2   rdf:type    ?Type1.
FILTER (?predicate1!=rdf:type)
FILTER (?predicate2!=rdf:type)         }

SELECT    ?Type1 ?predicate1 ?Type2
WHERE  {  ?s1   ?predicate1    ?s2.
          ?s1   rdf:type    ?Type1.
          optional{?s2   rdf:type    ?Type2.}
FILTER (?predicate1!=rdf:type)         }
```

Fig. 5. SPARQL statement to get the *PT* set

*5.1.3 Predicate-predicate Adjacent set (PP)*
Combining multiple semantic relation mappings is the core of building a basic graph pattern. However, not all combinations are reasonable, and we need to filter out inappropriate combinations via the predicate-predicate adjacent set *PP*, which is a part of *PT*. We can generate the *PP* set if we do not take the type in the *PT* set into consideration, denoted as $PP = \{(?-P_i/P_j)$ and $(P_i-?-P_j)\}$.

**Example 13**. Suppose we have the *PT* set (*dbp:knownFor-dbo:Book-dbo:publisher/ dbo:author*); we can then generate the *PP* set (*?-dbo:publisher/dbo:author*) and (*dbp:knownFor-?-dbo:publisher*).

*5.2 Semantic Relation Mapping*

*5.2.1 Phrase Mapping*

Relying on the paraphrase dictionary *ED*, the argument also can be mapped to the predicate such that we can obtain more and better candidate mappings and answer a query that cannot be answered by existing research studies. Table 3 shows an example of phrase mapping for query 1.

**Example 14**. For the query "*Give me cities in New Jersey with more than 100000 inhabitants*," there is a semantic relation <*cities, with, inhabitants*>. Because the *rel* "*with*" has no mapping that is predicate and "inhabitants" has no mapping that is a type, the existing methods cannot answer the query. In contrast, because *ED* contains the semantic equivalence between argument and predicate (i.e., "*inhabitants*"- "*dbo:populationTotal*"), we can get the semantic relation mapping <*dbo:City, dbo:populationTotal, ?inhabitant*> such that we can answer the query correctly.

Table 3. Phrase mapping

| Phrase | Predicate | Type |
|---|---|---|
| *books* | *dbo:awardedBook-0.5* | *dbo:Book-1.0* |
| *by* | | |
| *Kerouac* | | |
| *published* | *dbo:publisher-0.6* *dbo:publishedIn-0.6* *dbp:publishDate-0.6* | |
| *Viking_Press* | | |

*5.2.2 Semantic Relation Mapping*

In related research studies [28,29,31,32,33,34], they generate strict mapping for every phrase in a semantic relation so that they do not need to produce a semantic relation mapping by combination of phrase mappings. In contrast, to get better semantic relation mappings, our method of phrase mapping is relatively free (no affinity restriction) such that we need to combine these phrase mappings to get a semantic relation mapping via algorithm 2. Furthermore, there are many inappropriate semantic relation mappings by our method or related research studies, so we need to filter them out by *PT* in algorithm 2.

The basic idea of algorithm 2 contains four key points:

1) We select the appropriate mapping combination based on the adjacent relation between *type* and *predicate* in *PT*.

2) If *arg1* is mapped to a *predicate*, we will swap the position of *arg1* and *arg2* in the semantic relation mapping.

3) We recommend some candidate mappings to the phrase that has few mappings caused by spelling errors, and we can endure a few spelling mistakes by the Levenshtein distance.

4) Furthermore, we produce a subset of previous results (i.e., semantic relation mappings) after combination because the subset may be a correct semantic relation mapping while the superset is wrong. However, the subset must contain at least one determined argument.

---

**Algorithm 2** SRM (Semantic Relation Mapping)

**Require**: **Input**: Intention interpretation *I*
          Predicate-type adjacent set *PT*
   **Output**: Semantic relation mapping
   **Variables**: *Sm*, *Pm* and *Om* represent the mapping of *agr1*, *rel* and *agr2* respectively

1: Recommend some candidates mapping to semantic relation which have little mapping by *PT*
2: For each semantic relation do
3:   if($Sm \in \delta_{type}$)then
4:     if($Om \in \delta_{type}$&&(*Sm*,*Pm*,*Om*)satisfy *PT*) then output
5:     if(*Om* is null&&(*Sm*,*Pm*,*arg2*)satisfy *PT*)then output
6:     if($arg2 \in \delta_{constant}$)then output(*Sm*,?*x*,*arg2*)
7:     if($Om \in \delta_{predicate}$&&(*Sm*,*Pm*,*arg2*)satisfy *PT*)then output
8:   if($Sm \in \delta_{predicate}$)
9:     if($Om \in \delta_{type}$&&(*Om*,*Sm*,*arg1*) satisfy *PT*) then output
10:     if(*Om* is null)then output(*arg2*,*Sm*,*arg1*)

11: if($Om \in \delta_{type}$ && ($arg1,Pm,Om$) satisfy $PT$) then output
12: if($Om$ is null) then output($arg1,Pm,arg2$)
13: if($Om \in \delta_{predicate}$) then output($arg1,Om,arg2$)
14: Produce the subset of all above results and output
15: Remove all duplicate semantic relation mapping

There are four examples corresponding to the four key points in algorithm 2 as follows:

**Example 15**. Consider $PT$={(*dbo:Book-dbo:publisher/dbo:author/dbo:publishedIn*)} and the semantic relation <*books, published, Viking_Press*>; if "*books*" is mapped to "*dbo:Book*," "*published*" can only be mapped to "*dbo:publisher*" or "*dbo:publishedIn*" rather than "*dbp:publishDate*" in Table 3. Thus, we can get two semantic relation mappings as shown in Table 4 (i.e., <*dbo:Book, dbo:publisher,Viking_Press-2.6*> and <*dbo:Book, dbo:publishedIn, Viking_Press-2.6*>). Moreover, we cannot discard other predicates that may be the right mapping, so we get another semantic relation mapping (i.e., <*books, dbp:publishDate, Viking_Press*>), which does not contain "*dbo:Book*."

**Example 16**. The argument "*books*" is also mapped to the predicate "*dbo:awardedBook*," so we adjust "*dbo:awardedBook*" as the predicate and swap the position of *arg1* and *arg2*. Then, we get a semantic relation mapping < *Viking_Press, dbo:awardedBook, books* >.

**Example 17**. Due to the *rel* "*by*" having no mapping and the argument "*books*" being mapped to "*dbo:Book*," we recommend some candidate mappings to "*by*" that rely on $PT$ = (*dbo:Book-dbo:publisher/dbo:author/ dbo:publishedIn*). However, even if the Levenshtein Distance is used, the *rel* "*by*" still has no mappings because it is not caused by spelling errors.

**Example 18**. Finally, we generate the mapping <*dbo:Book, ?y, Viking_Press*>, which is a subset of <*dbo:Book, dbo:publishedIn, Viking_Press*>. Because the predicate "*dbo:publishedIn*" is wrong according to the RDF data in Fig. 1, the subset will be very useful for answering the query.

Finally, we can get the semantic relation mappings for Query 1 as shown in Table 4.

Table 4. Semantic relation mapping

| R | Mapping | Subset |
|---|---|---|
| R1<*books, by, Kerouac*> | *dbo:Book, ?X, Kerouac -2.0*<br>*Kerouac, dbo:awardedbook, books -1.5* | *Books, ?X, kerouac -1.0*<br>*Kerouac, ?X, books -1.0* |
| R2<*books, published, Viking_press*> | *dbo:Book, dbo:publisher,Viking_Press -2.6*<br>*dbo:Book, dbo:publishedIn, Viking_Press -2.6*<br>*books, dbp:publishDate, Viking_Press -1.6*<br>*Viking_Press, dbo:awardedbook, books -1.5* | *dbo:Book, ?y, Viking_Press -2.0*<br>*books, ?y, Viking_Press -1.0*<br>*Viking_Press, ?y, books -1.0* |

*5.3 Building Basic Graph Patterns*

In related research studies [28,29,31,32,33,34], they select semantic relation mappings with higher scores to form the basic graph pattern, but there is a large number of inappropriate mapping combinations. We filter out inappropriate predicate-predicate combinations by *PP* (section 5.1.3) and delete irrational basic graph patterns that do not satisfy certain rules.

*5.3.1 Rules of Basic Graph Patterns*

There is an irrational basic graph pattern in a few cases. We need to delete it from $\delta_G$ if it does not satisfy one of the following:
● All question items must appear in the basic graph pattern.
● All aggregate items must appear in the basic graph pattern.

*5.3.2 Building Basic Graph Patterns*

A group of mappings of all semantic relations is collected together to form a BGP (basic graph pattern). To get a BGP, we need an algorithm to combine multiple semantic relation mappings. Moreover, there is mismatch between one semantic relation mapping and others such that we need to filter them out. Therefore, we propose algorithm 3, and its basic idea is as follows: 1) we use a recursive method to get top-*k* basic graph patterns with the highest scores; and 2) we select the appropriate matching between one semantic relation mapping and another one by *PP*.

In algorithm 3, $\delta_i$ represents all mappings of $R_i$, $pp(G+m)$ represents the adjacent relationship between predicate *p* (which is the predicate in the candidate semantic relation mapping *m*) and predicate (*p*'s adjacent predicates in *G*), and *score*(*G+m*) represents the score of the candidate basic graph pattern *G+m*.

**Algorithm 3** BBGP (Building Basic Graph Pattern)
**Require**: **Input**: Semantic relation number $n$
　　　　　Semantic relation mapping $\delta=\{\delta_1, ..., \delta_n\}$
　　　　　Predicate-type adjacent set $PT$
**Output**: Basic graph pattern set $\delta_G$
1: Get predicate-predicate adjacent set $PP$ from $PT$
2: For each semantic relation mapping $\delta_i$ in $\delta$
3: 　　In order of score of each mapping $m$ in $\delta_i$
4: $k=1$, //$k$-th semantic relation is processing
5: $G=\emptyset$,//temporary store partial basic graph pattern
6: $\delta_G=\emptyset$//store top-$k$ basic graph pattern
7: Recursive($PP,\delta,k,G,n,\delta_G$)
**Recursive**($PP, \delta,k,G, \delta_G$)
1: if($k==n$)
2: 　　for each semantic relation mapping $m$ in $\delta_k$
3: 　　　　if ($pp(G+m) \in PP \&\& (G+m)$ satisfy rules)
4: 　　　　　　if $score(G+m)>$min_score($\delta_G$)
5: 　　　　　　　　update $\delta_G$ by $G+m$
6: 　　　　　　Else return;
7: if($k<n$)
8: 　　for each mapping m in $\delta_k$
9: 　　　　if ($pp(G+m) \in PP$)
10: 　　　　　Recursive($PP, \delta,k+1,G+m, \delta_G$)

**Example 19**. For query 1, because the predicate "*dbo:publisher*" has the *PP* set (?-*dbo:publisher/dbo:author*), we do not combine the semantic relations (*Kerouac, dbo:awardedBook, books*) and (*dbo:Book, dbo:publisher, Viking_Press*) because "*dbo:awardedBook*" is not adjacent to "*dbo:publisher*." Table 5 shows some basic graph patterns for query 1.

Table 5. Basic graph pattern

|  | Rm1+Rm2 | Score |
|---|---|---|
| *BGP1* | *<dbo:Book, ?X, Kerouac-2.0> <dbo:Book, dbo:publisher,Viking_Press-2.6>* | *5.2* |
| *BGP2* | *<dbo:Book, ?X, Kerouac-2.0> <dbo:Book, dbo:publishedin, Viking_Press-2.6>* | *5.2* |
| *BGP3* | *<dbo:Book, ?X, Kerouac-2.0> <dbo:Book, ?Y, Viking_Press-2.0>* | *4.0* |
| ... | …… | ... |

# 6. Translation

*6.1 Offline*

For the query "*How many student in Classx?*," consider two data formats: 1) *<dbr:Classx, dbo:studentNum, 2>*; 2) *<dbr:Classx, dbo:studentName, name1><dbr:Classx, dbo:studentName, name2>*. For the above two data formats, we need to use a different aggregate category SUM (or null for the first case) or COUNT (for the second case). On the one hand, we do not know the data format from the query. On the other hand, even if many queries have the same sentence structure, different aggregate items may correspond to different data formats. Thus, we cannot assign SUM or COUNT arbitrarily. However, if we can confirm whether the aggregate item $a_i$ is a numeric value or not, everything becomes simpler for aggregate categories such as *sum*, *count*, $<$, $>$, *max*, *min* and *avg*. Upon careful study, we find that $a_i$ is a numeric value only if $a_i$ is *arg2* in a semantic relation and the predicate belongs to a specific predicate set $\delta_{numeric}$. The SPARQL to get the specific predicate set $\delta_{numeric}$ is shown in Fig. 6. The DBpedia data of the online query web (http://dbpedia.org/sparql/) is up-to-date as of April 2016, and the predicate, which has a data type in the format "*day, kilometer….,*" has been replaced so that all specific predicates are included in $\delta_{numeric}$.

```
select distinct ?numeric_predicate
where {?numeric_predicate   rdfs:range   ?y.
values ?y{
        xsd:double
        xsd:float
        xsd:nonNegativeInteger
        xsd:positiveInteger
        xsd:integer}         }
```

Fig.6. The SPARQL to get the numeric predicates set $\delta_{numeric}$

*6.2 Translate Basic Graph Patterns*

For basic graph patterns, there are two parts that need to be converted. 1) If the mapping of *arg1* (or *arg2*) is a *type*, we need to construct a new triple that represents the relationship between *arg1* (or *arg2*) and the *type* and then transform the mapping to *arg1* (or *arg2*). However, we must avoid generating duplicate triples. 2) If the argument is a variable, we need to add '?' before the argument to make it a question node.

**Example 20**. There is a basic graph pattern {<dbo:Book ?x Kerouac>,<dbo:Book dbo:publisher Viking_Press>}. According to the above rules, we generate SPARQL statements as follows:
*?books rdf:type dbo:Book.*
*?books ?x Kerouac.*
*?books dbo:publisher Viking_Press.*

*6.3 Translate Aggregation (TA)*

In this part, we have shown the algorithm TA (Algorithm 4), which translates aggregation into SPARQL. There are two points that need to be explained: 1) due to the complexity of aggregation, we divided it into four levels, which can include most aggregate categories except for nested queries; and 2) due to the diversity of basic graph patterns, the aggregation cannot be suitable for all basic graph patterns, and we need to translate aggregation for each basic graph pattern.

**Example 21**. For the query "*What is the largest city in Australia?*," we can get the intention interpretation *I*={*S*={<*city, in, Australia*><*city, largest, #*>}, *Q*={*city*}, *A*={*city, largest*}} (see example 10 in section 4.3). After the basic graph pattern is translated, we may get the set that contains the SPARQL statement of the basic graph pattern, question item and aggregation (i.e., {{<*Australia, dbo:largestCity, ?city*>}, {*city*}, ∅} and {{<*?city rdf:type dbo:City*><*?city ?x Australia*><*?city dbo:populationTotal ?x1*>}, {*city*}, {*?x1,max*}}). We need to translate the aggregation using *primary_aggregation*() for the first case and the second one using *higher_numeric_ aggregation*(). Furthermore, the aggregate item has been replaced by ?x1 due to the semantic relation mapping having changed in the second basic graph pattern.

---

**Algorithm 4** TA(Translate Aggregation)
**Require**: **Input**: Basic graph pattern set $\delta_G$
         Aggregate item and category set *A*
         Question item set *Q*
         Predicate set $\delta_{numeric}$ whose *arg2* is numeric
    **Output**: SPARQL with aggregation
1: For each $\delta$ in $\delta_G$ do
2:    If *A* is primary_ aggregation
3:        primary_ aggregation()
4:    Else If (question item==aggregate item)
5:        intermediate_ aggregation()
6:    Else If aggregate item is *agr2* and predicate∈ $\delta_{numeric}$
7:        higher_numeric_ aggregation()
8:    Else higher_nonnumeric_ aggregation()
9: Add select, where and distinct clause
**Procedure primary_aggregation()**
10:    If *agr1* or *arg2* in $\delta$ has constant *y* then
11:        Replace *y* to ?*x* and Add "FILTER regex(?*x*, "*y*")"
12:    If aggregate category=="same" then

|  |
|---|
| Replace a to ?*x* and ?*y* respectively and Add "FILTER (?*x*=?*y*)" |
| 13:   If aggregate category in predicates set of δ then |
| 14:       Don't do anything |
| **Procedure intermediate_aggregation()** |
| 15:   If aggregate category ==avg/max/min then |
| 16:       Add "avg/max/min(?*x*)" |
| 17:   If aggregate category ==count or sum then |
| 18:       If aggregate item *x* is *arg2* and predicate $\in \delta_{numeric}$ |
| 19:           Add "sum(?*x*)" |
| 20:       Else Add "count(?*x*)" |
| **Procedure higher_numeric_aggregation()** |
| 21:   If aggregate category== >/< *a* then |
| 22:       Add "FILTER(?*x*>/<*a*)" |
| 23:   If aggregate category==max/min |
| 24:       Add "order by DESC/ASC(aggregate item) Limit 1" |
| 25:       If there are aggregate category==first/second….. |
| 26:           Add "offset 1/2…." |
| **Procedure higher_nonnumeric_aggregation()** |
| 27:   If aggregate category== >/< *a* then |
| 28:       Add "group by question item having(count(?*x*)>/<*a*) |

# 7. Experiments

QALD (http://qald.sebastianwalter.org/) is a benchmark for RDF Q/A questions. We produce top-10 SPARQL statements and execute them on the Virtuoso SPARQL Query Editor (http://dbpedia. org/sparql/). If there is one of the top 10 that can answer the question correctly, we deem that we can answer the question.

*7.1 Question Dataset*

In our experiments, we use three sets of question datasets from QALD: 1) QALD-3 testing questions, 2) QALD-3 training questions and 3) aggregate questions extracted from both of them.

To distinguish the ID of testing questions and training questions, "(train)" means that the question comes from the QALD-3 training question set, and other questions come from the QALD-3 testing question set.

*7.2 Comparison*

*7.2.1 Component Comparison*

*Query Understand Comparison.* In our method, we can identify more semantic relations, which is often overlooked by existing research studies. The primary reason is that existing methods for identifying semantic relations rely too much on the verb phrase (more rigorous discussion is available in section 1). In contrast, we identify semantic relations relying only on dependency structures that existing methods also use. In other words, in the process of transforming dependency structures to semantic relations, we consider more regarding the dependency relationship among phrases and reduce the step to identify the verb phrase in question and paraphrase dictionary *D*. We have compared the number of questions that cannot be understood correctly, as shown in Table 6.

Table 6. Query understand comparison

|  | Graphdata[31] | NLAQ |
|---|---|---|
| Error semantic relation | 14 | 3 |
| Can't identify aggregation | 22 | 0 |

*Mapping Comparison* (*ED*). We propose the extended paraphrase dictionary *ED*, which maps arguments of the semantic relation to predicates. Existing research studies cannot get predicate mappings for semantic relations that do not contain a verb phrase, and the extended *ED* may help us solve this problem (see example 14 in section 5.2). There are some questions that benefit from the extended *ED*, as shown in Table 7.

Table 7. Some aggregate questions that benefit from the extended *ED*

| Give me all books by William Goldman with more than 300 pages. |
|---|
| *Do Prince Harry and Prince William have the same mother?* |
| *Which state of the USA has the highest population density?* |
| *Which countries have more than two official languages?* |
| *Give me the websites of companies with more than 500000 employees.* |
| *Which caves have more than 3 entrances?* |

*Mapping Comparison* (*PT*). To reduce inappropriate combinations, we propose the predicate-type adjacent set *PT* (and the subset *PP* of *PT*) to filter out inappropriate combinations in semantic relation mappings (and basic graph patterns). Many factors (such as different RDF data, different questions) will lead to a different filter ratio of *PT*. Therefore, we use an example to illustrate the effectiveness of *PT*. For the query "*Who produced the most films?,*" we have a standard SPARQL statement as follows:

*SELECT DISTINCT ?person*
*WHERE { ?film rdf:type dbo:Film .*
*?film dbp:producer ?person .*
*?person rdf:type dbo:Person }*
*ORDER BY DESC(COUNT(?film))*
*LIMIT 1*

There is no doubt that all methods can map "*produced*" to many predicates, such as "*dbp:producer*," "*dbp:producedBy*," "*dbp:coProducer*" and so on. However, a further obstacle has been presented. According to statistics, there are **182** predicates that contain the string "*produce*" in the RDF data set (DBpedia). Existing methods depend on similarity scores alone to select candidate mappings, and it is hard to get the suitable predicate. In contrast, our method can filter out many inappropriate candidate mappings. Firstly, we get two subsets of *PT* (i.e., the adjacent predicate set of "*dbo:Film*" and "*dbo:Person*"), denoted as A1:{*dbo:Film-predicate1/…*} and A2:{*predicate'1/…-dbo:Person*}. Secondly, based on A1, we can filter out predicates that are not adjacent to *dbo:Film* so that there are **64** predicates left. Thirdly, similarly, there are **21** predicates left by using the adjacent set A2. Finally, compared with the result of existing research studies, we will get very suitable candidate predicates that rely on similarity scores. Furthermore, if there are multiple semantic relations in the query, we can continue to filter out some predicates by determining whether two predicates in two semantic relations are neighbors in *PP*.

Table 8. The ability to filter by *PT*

| | |
|---|---|
| number of predicates which contain strings "produce" | 182 |
| number of predicates which is adjacent to "Film" | 64 |
| number of predicates which is adjacent to "Film" and "Person" | 21 |

*7.2.2 Algorithm Comparison*

We also compare NLAQ to other systems that can answer some aggregate queries as shown in Table 9. The number of questions that can be answered is only the number of correct answers (i.e., top-*k* set including one correct SPARQL statement that returns the desired answer), and the statistics are derived from the QALD-3 evaluation results. We only compare the aggregate questions from the QALD-3 testing questions that most algorithms deal with. Although the best system is squall2sparql [28], which can answer 20 aggregate questions, the input of squall2sparql is controlled English questions rather than real natural language questions. For the query "*Give me all world heritage sites designated within the past five years.,*" the input of squall2sparql is "*Give me all WorldHeritageSite whose dbp:year is between 2008 and 2013.*" As shown in Table 9, NLAQ is obviously better than the other methods.

Table 9. Comparison of several algorithms

| Algorithm | Number | Test-Questions ID |
|---|---|---|
| Squall2sparql | 20 | 4,5,11,12,13,15,23,25,26,32,38,50,61,68,73,80,85,86,88,99 |
| NLAQ | 14 | 4,5,15,23,26,38,59,61,68,73,80,85,86,92 |
| CASIA | 6 | 4,26,68,85,86,93 |
| Scalewelis | 6 | 4,23,32,50,68,85 |
| RTV | 6 | 26,32,38,68,73,86 |
| SWIP | 5 | 38,68,85,86,88 |
| Intui2 | 4 | 38,68,85,86 |
| Template | 4(train) | 58(train),69(train),88(train),92(train) |
| Graphdata[31] | 0 | —— |

Furthermore, to show the superiority of our method, we contrast another aspect of it. There are 99 natural language questions in the QALD-3 testing questions, which include general questions and aggregate questions. The best system that can answer most questions is the graph data-driven approach [31] except squall2sparql [28], and it can answer 32 questions correctly. Its correct answer rate is 32.32% (32/99 general/aggregate questions), lower than our answer rate of 68.75% (33/48 aggregate questions). As known to all, the aggregate questions are harder to answer than general questions. Thus, we can conclude that our method is very effective.

Table 10. Accuracy comparison

|  | methods | Right | rate |
|---|---|---|---|
| Aggregate query (48) | NLAQ | 33 | 0.68 |
| General query (99) | Squall2sparql | 77 | 0.77 |
|  | Graphdata[31] | 32 | 0.32 |
|  | RTV | 30 | 0.30 |
|  | CASIA | 29 | 0.29 |
|  | Intui2 | 28 | 0.28 |
|  | DEANNA | 21 | 0.21 |
|  | SWIP | 14 | 0.14 |
|  | Scalewelis | 1 | 0.01 |

*7.3 Effectiveness Evaluation*

We use aggregate questions from the QALD-3 training questions and testing questions in our experiments. We can answer 33 questions correctly in all 48 aggregate questions. We show the experimental results in Table 11.

Table 11. All aggregate questions in QALD-3

| **Can Answer by NLAQ (33)** ||
|---|---|
| ID | **From Testing Questions (14)** |
| 4 | *How many students does the Free University in Amsterdam have?* |
| 5 | *What is the second highest mountain on Earth?* |
| 15 | *What is the longest river?* |
| 23 | *Do Prince Harry and Prince William have the same mother?* |
| 26 | *How many official languages are spoken on the Seychelles?* |
| 38 | *How many inhabitants does Maribor have?* |
| 59 | *Which U.S. states are in the same timezone as Utah?* |
| 61 | *How many space missions have there been?* |
| 68 | *How many employees does Google have?* |
| 73 | *How many children did Benjamin Franklin have?* |
| 80 | *Give me all books by William Goldman with more than 300 pages.* |
| 85 | *How many people live in the capital of Australia?* |
| 86 | *What is the largest city in Australia?* |
| 92 | *Show me all songs from Bruce Springsteen released between 1980 and 1990.* |
| ID | **From Training questions (19)** |
| 11 | *Which countries have places with more than two caves?* |
| 17 | *Give me all cities in New Jersey with more than 100000 inhabitants.* |
| 20 | *How many employees does IBM have?* |
| 24 | *Which mountain is the highest after the Annapurna?* |
| 26 | *Which bridges are of the same type as the Manhattan Bridge?* |
| 30 | *Which state of the USA has the highest population density?* |
| 34 | *Which countries have more than two official languages?* |
| 40 | *What is the highest mountain in Australia?* |
| 47 | *What is the highest place of Karakoram?* |
| 52 | *Which presidents were born in 1945?* |
| 58 | *Who produced the most films?* |
| 61 | *Which mountains are higher than the Nanga Parbat?* |
| 67 | *Give me the websites of companies with more than 500000 employees.* |
| 69 | *Which caves have more than 3 entrances?* |
| 76 | *How many films did Hal Roach produce?* |
| 81 | *Which country has the most official languages?* |

| 88 | *How many films did Leonardo DiCaprio star in?* |
|---|---|
| 91 | *Which organizations were founded in 1950?* |
| 92 | *What is the highest mountain?* |
| | **Can't Answer by NLAQ (15)** |
| ID | **From Testing Questions (12)** |
| 1 | *Which German cities have more than 250000 inhabitants?* |
| 11 | *Who is the Formula 1 race driver with the most races?* |
| 12 | *Give me all world heritage sites designated within the past five years.* |
| 13 | *Who is the youngest player in the Premier League?* |
| 16 | *Does the new Battlestar Galactica series have more episodes than the old one?* |
| 25 | *Which U.S. state has been admitted latest?* |
| 32 | *How often did Nicole Kidman marry?* |
| 50 | *Was the Cuban Missile Crisis earlier than the Bay of Pigs Invasion?* |
| 75 | *Which daughters of British earls died in the same place they were born in?* |
| 88 | *Which films starring Clint Eastwood did he direct himself?* |
| 93 | *Which movies did Kurosawa direct after Rashomon?* |
| 99 | *For which label did Elvis record his first album?* |
| ID | **From Training Questions (3)** |
| 5 | *How many monarchical countries are there in Europe?* |
| 19 | *Is Egypts largest city also its capital?* |
| 46 | *Is Frank Herbert still alive?* |

Moreover, because the answer may be precomputed and stored as an attribute in DBpedia, there are 8 aggregate questions that can be answered by triples rather than an aggregate function. For this case, we can correctly answer all of them. For example, for query ID=68 "*How many employees does Google have?*" there is the triple <*Google, dbo:numberOfEmployees, 500000*> in DBpedia such that we have no use for the aggregate function COUNT.

Table 12. The predicate that contains aggregation

| ID | Questions | Predicate |
|---|---|---|
| 4 | *How many students does the Free University in Amsterdam have?* | *dbo:numberOfStudents* |
| 38 | *How many inhabitants does Maribor have?* | *dbp:populationTotal* |
| 68 | *How many employees does Google have?* | *dbo:numberOfEmployees* |
| 85 | *How many people live in the capital of Australia?* | *dbp:populationTotal* |
| 86 | *What is the largest city in Australia?* | *dbo:largestCity* |
| 20(train) | *How many employees does IBM have?* | *dbo:numberOfEmployees* |
| 47(train) | *What is the highest place of Karakoram?* | *dbo:highestPlace* |

*7.4 Causal Analysis*

There are 15 questions that we cannot answer. As shown in Table 13, the main reasons are:

1) There are incorrect dependency structures that come from the Stanford Parser. For the query "*Was the Cuban Missile Crisis earlier than the Bay of Pigs Invasion?*" there are two wrong dependency structures: "*amod(Crisis-5, Cuban-3)*" and "*dep(Invasion-12, Pigs-11).*" We cannot get two constants, "*Cuban Missile Crisis*" and "*Bay of Pigs Invasion,*" so we cannot answer this query.

2) There is implicit information contained in a question. For the query "*Give me all world heritage sites designated within the past five years,*" our method cannot understand "*within past five years.*"

3) We cannot find the semantic relation. For the query "*How many monarchical countries are there in Europe?,*" we cannot find the semantic relation <*countries,#,monarchical*> because the dependency structure "*amod(countries-4, monarchical-3)*" does not belong to the dependency structure set of the semantic relations.

4) We cannot find a mapping for the phrase. For the query "*Is Frank Herbert still alive?,*" we cannot find the mapping "*dbo:deathDate*" for the phrase "*alive*" in the semantic relation <*Frank_Herbert, alive, #*>.

Table 13. Classification of causes

| Error result of Stanford parser | 11,50,75,93,99 |
|---|---|
| Implicit Information | 12,16,32,88,19(train), |
| Missing semantic relation | 1,25,5(train) |
| No mapping | 13,46(train) |

# 8. Other challenges

*8.1 Top-K*

On the one hand, sometimes a query may have multiple corresponding BGPs (Basic Graph Patterns) with identical scores. If there are multiple BGPs in the top-*k* set with an identical lowest score, we arbitrarily break the tie of *k* and accept all these BGPs. On the other hand, if multiple BGPs are only different in namespace, we regard them as one in the top-*k* set. For example, there are two basic graph patterns {<*Maribor dbo:populationTotal inhabitants*>} and {<*Maribor dbp:populationTotal inhabitants*>}, and we regard them as one in the top-*k* set.

*8.2 Union Pattern*

Due to the complexity of aggregation, a union pattern cannot be used arbitrarily.

For the query "*How many inhabitants does Maribor have?*" we can get the basic graph pattern {<*Maribor dbo:populationTotal inhabitants*>} and another similar basic graph pattern {<*Maribor dbp:populationTotal inhabitants*>}. According to the universal rule, we will combine them together and then translate the result to a SPARQL statement as follows:

*Select sum(?inhabitants)*
*Where { { ?x dbo:populationTotal ?inhabitants}*
*UNION { ?x dbp:populationTotal ?inhabitants }*
*FILTER regex(?x, "Maribor") }*

However, the above-mentioned SPARQL statement is incorrect, because two namespaces have the same number of inhabitants. As a result, we will get twice as many as the number of inhabitants.

In contrast, for the query "*Which organizations were founded in 1950?*" we should use "UNION." We can obtain the SPARQL statement as follows:

*SELECT DISTINCT ?uri*
*WHERE { ?uri rdf:type dbo:Organisation .*
*{?uri dbo:formationYear ?date . }*
*UNION { ?uri dbo:foundingYear ?date. }*
*UNION { ?uri dbp:foundation ?date . }*
*UNION { ?uri dbp:formation ?date . }*
*FILTER regex(?date,'^1950') . }*

To solve the above problems, we design one rule: if the "UNION" pattern contains a numeric variable that has the aggregate category "SUM," we split the "UNION" pattern into multiple SPARQL statements. The others do not need to be addressed anymore because "DISTINCT" will solve the problem.

*8.3 PT Set and PP Set*

Consider $PT = \{(T_i - P_k - T_j) \text{ and } (P_i - T_k - P_j)\}$ and its subset $PP = \{(? - P_i/P_j) \text{ and } (P_i - ? - P_j)\}$; as we can see, the two sets are just related to two connected triples, so we define the path length of this set as two. This is because most queries involve two connected triples in our question set. If most queries involve more connected triples in another application environment, the *PT* set and *PP* set can still be useful, and we only need to increase the path length of *PT*. However, a longer path length will lead to a larger size of the *PT* set and *PP* set.

*8.4 Namespace*

There are a few types that have the predicate "*dbo:type/…*" instead of "*rdf:type,*" such as "*dbr:China_Aid dbo:type dbr:Nonprofit_organization*." We have recorded these types so that we can translate the basic graph pattern into the correct SPARQL statement.

*8.5 Levenshtein Distance*

Natural language questions contain various phrases. For the query "*Give me all cities….,*" we must identify that "*cities*" should be mapped to "*city*." Because words have tenses and changes of tenses appear on the right side of the word, if the word has no mapping, we will relax the restriction of the Levenshtein Distance and allow three letters on the right side to be different. In general, we allow one letter at any location to be different. Thus, we can accommodate a few spelling mistakes.

# 9. Conclusions and Future Work

We have made a first step toward processing natural language aggregate queries over RDF data without restrictions. Although there is some literature that can only answer a small number of aggregate queries over RDF data, they have some limitations (i.e., controlled English questions, interactive information and query template). We propose a framework called NLAQ that can automatically identify the aggregation (AIII algorithm) and transform it into a SPARQL aggregate statement (TA algorithm). Moreover, the TA algorithm can effectively distinguish numeric aggregate items, which will greatly affect the aggregate result.

Compared with existing studies, we can identify semantic relations much more effectively. Existing research studies regarding the identification of semantic relations entirely depend on the verb phrase in the query and the paraphrase dictionary *D*, which records the semantic equivalence between verb phrases and predicates. Therefore, they can identify the triples whose relation phrase is a verb phrase and overlook other triples. We propose an algorithm called AIII that considers more about the dependency relationships among phrases rather than identifying the verb phrase so that we can avoid missing triples whose predicates are not verb phrases.

We propose the extended paraphrase dictionary *ED* and predicate-type adjacent set *PT* to yield better candidate mapping. Compared with existing studies, we do not simply map the relation phrase to the predicate and filter out mappings by similarity score. During the mapping stage, on the one hand, to get more candidate mappings, we propose the extended paraphrase dictionary *ED*, which adds the semantic equivalence between arguments of the semantic relation and predicates to the existing paraphrase dictionary *D*. On the other hand, we propose the predicate-type adjacent set *PT* and the subset *PP* of *PT* to filter out inappropriate mapping combinations in semantic relations and basic graph patterns, respectively. In summary, *ED* improves the semantic relation mapping, while *PT* improves the semantic relation mappings and basic graph patterns, so that we can answer more queries.

Overall, NLAQ not only can answer aggregate queries over RDF data but also can improve natural language queries, such as by identifying better semantic relations and filtering out inappropriate mapping combinations in semantic relations and basic graph patterns.

There are some related issues that are worth studying in the future. Firstly, how to answer an implicit query, such as "……than the old one?," is a very valuable issue. Secondly, our method depends on the dependency structure that comes from the Stanford Parser. If there is a way to increase the accuracy of the dependency structure, we can answer more questions. Thirdly, queries that require nested SPARQL statements are worth exploring.

# Acknowledgment

This work was supported by the National Natural Science Foundation of China (Project No. 60940032, No. 61073034 and No. 61370064) and the Program for New Century Excellent Talents in University of Ministry of Education of China under Grant No. NCET-10–0239.